\documentclass[letterpaper, 10pt, journal, twoside]{IEEEtran}
\usepackage{graphicx}      
\usepackage{cite}        

\usepackage{amsmath,amssymb,epsfig}
\usepackage{enumerate,url,wasysym,balance}
\usepackage{multirow}
\usepackage{xcolor}

\newcommand \mcU{\mathcal{U}}
\newcommand \mcX{\mathcal{X}}

\newcommand \ba{\mathbf{a}}
\newcommand \bc{\mathbf{c}}
\newcommand \bff{\mathbf{f}}
\newcommand \bg{\mathbf{g}}
\newcommand \br{\mathbf{r}}
\newcommand \bu{\mathbf{u}}

\newcommand \bx{\mathbf{x}}

\begin{document}

\title{Advocating Feedback Control for Human-Earth System Applications}


\author{Guido Cavraro
\thanks{This work was authored by the National Renewable Energy Laboratory, operated by Alliance for Sustainable Energy, LLC, for the U.S. Department of Energy (DOE) under Contract No. DE-AC36-08GO28308. Funding provided by the NREL Laboratory Directed Research and Development Program. The views expressed in the article do not necessarily represent the views of the DOE or the U.S. Government. The U.S. Government retains and the publisher, by accepting the article for publication, acknowledges that the U.S. Government retains a nonexclusive, paid-up, irrevocable, worldwide license to publish or reproduce the published form of this work, or allow others to do so, for U.S. Government purposes.}
\thanks{{\tt guido.cavraro@nrel.gov}.}
}

\maketitle

\begin{abstract}
This paper proposes a feedback control perspective for Human-Earth Systems (HESs) which essentially are complex systems that capture the interactions between humans and nature.
Recent attention in HES research has been directed towards devising strategies for climate change mitigation and adaptation, aimed at achieving environmental and societal objectives. However, existing approaches heavily rely on HES models, which inherently suffer from inaccuracies due to the complexity of the system. Moreover, overly detailed models often prove impractical for optimization tasks.
We propose a framework inheriting from feedback control strategies the robustness against model errors, because inaccuracies are mitigated using measurements retrieved from the field. The framework comprises two nested control loops. The outer loop computes the optimal inputs to the HES, which are then implemented by actuators controlled in the inner loop.
Potential fields of applications are also identified. 
\end{abstract}

\section{Introduction}

\begin{figure}[tb]
\centering
\includegraphics[width=0.92\columnwidth]{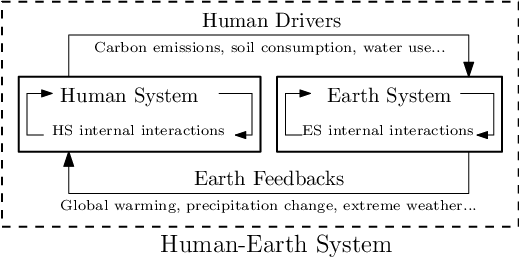}
\caption{Schematic representation of the HES. The Human and Earth system blocks feature internal interactions and are linked by feedback.}
    \label{fig:HES}
\end{figure}

A significant effort has been dedicated to understand how climate changes because of human actions and how to limit the global warming, e.g., through decarbonization paths and policies to. This field is extremely challenging  because it lies in the intersection of climate science, economics, engineering, and social science.
Classically, research on human systems and the Earth system was conducted separately, even though humans and the environments naturally interact and form a single complex system referred to as HES~\cite{calvin2018integrated}, see Figure~\ref{fig:HES}.

Human systems encompass various aspects of society, e.g., economic and social dynamics.
Earth systems typically describe the interaction between the climate and the global scale biogeochemistry.
Even though Earth science has always recognized that humans are an important component, key characteristics of the Anthropocene, e.g., human agency and social and economic networks, have not been dynamically represented in classic Earth systems.
Capturing these dynamics in a new generation of Earth system models allow to address a number of critical questions about socio-ecological interactions~\cite{donges2017closing}.
HES science adopt a holistic approach to capture interactions and \emph{feedbacks} within and between human systems yielding advantages such as the ability to represent ecosystem dynamics more realistically at longer timescales, and provide insights that cannot be generated using only Earth  models.
Surprisingly, the literature on HES lacks substantial contributions from control engineers, who possess expertise in modeling, handling, and controlling complex dynamical systems.

Scientists developed several \emph{integrated assessment models} (IAMs), i.e.,  quantitative models that describe the global interplay between climate, economy, production sectors, and society. 
IAMs integrate theories and empirical results from several disciplines to  understand how human development and societal choices affect each other and the natural world, including climate change~\cite{hope2020integrated}. IAMs constitute a fundamental tool for the development and validation of decarbonization paths.
This line of research has been so impactful that Prof. W. Nordhaus, who developed one of the first IAMs (DICE), was awarded the Nobel Prize.
As we will show next, IAMs act in an \emph{open loop} fashion and provide  results affected by model inaccuracies.
This rises questions about straightforwardly applying IAMs' results in practical applications~\cite{pindyck2017use}. Section~\ref{sec:IAM} reports a brief discussion about IAMs.

Since feedback control features well-established properties of errors and uncertainty rejection, we argue that a \emph{closed loop} approach can be useful to overcome the limitations of IAMs.
This work then suggests a feedback control perspective for studying HESs. Precisely, in Section~\ref{sec:feedback}, we propose a control framework consisting of two loops.
Section~\ref{sec:opt} describes the outer control loop whose goal is to obtain the optimal control inputs steering the HES system toward desirable states, i.e., states meeting properties like a bounded average temperature increase or a minimum global welfare level.
Section~\ref{sec:act} proposes an inner control loop whose goal is to ensure that the aforesaid optimal setpoints are implemented in the HES    .

\textit{Notation}. Vectors are denoted by lower case boldface letters, respectively. 
Calligraphic symbols denote sets; $\mathbb{R}$ denotes the set of real and complex numbers. 



\section{IAMs and the current approach}
\label{sec:IAM}

We can broadly group the IAMs into simple and complex IAMs.
The first ones rely on simplified equations and are often used to calculate the \emph{social cost of carbon}, a measure of the quantifiable costs and benefits of emitting one additional ton of $\text{CO}_2$. 
Among them, we mention DICE, PAGE, and FUND. 
DICE is mainly used to compute optimal emissions pathways and the social cost of carbon~\cite{barrage2023policies}.
PAGE includes regional details and expresses all the main inputs as probability distributions so that the risk of climate change could more easily be appreciated~\cite{hope1993policy}.
FUND study the role of international capital transfer in climate policies
and features more regional and sectorial detail~\cite{anthoff2014climate}.

In contrast, complex IAMs like GCAM~\cite{calvin2019gcam}, IMAGE~\cite{bouwman2006integrated}, WITCH~\cite{de20092008}, MERGE~\cite{manne1995merge}, and EPPA~\cite{paltsev2005emissions}, are built on interconnected modules representing various aspects of HESs, including the global economy, energy systems, land usage, climate, and carbon cycles. 
Depending on the purpose of the IAM, the corresponding climate and carbon modules vary in complexity and resolution, e.g., see MAGICC~\cite{meinshausen2011emulating} and
HECTOR~\cite{hartin2015simple}. 

Carbon cycle simulators essentially relate carbon emissions from anthropic activities and  natural processes, carbon absorption from soil, vegetation, and ocean, radiating forcing, and temperature.
Of particular interest is how they model \emph{radiative forcing}, i.e., the change in net radiation balance (solar plus long-wave
) of the surface-troposphere system.
Carbon cycle simulators typically express radiative forcing as a function of atmospheric concentrations of greenhouse gases (GHGs), e.g., $\text{CO}_2$, Ozone, Methane, Nitrus oxide.
Even though climate models in different IAMS may consider different spatial or time scales, in principle they model the global carbon cycle and compute future concentrations of GHGs from given emissions, calculate the global mean radiative forcing from GHG concentrations, and convert the radiative forcing to global mean temperature~\cite{hartin2015simple}.
%


Climate engineering, or geoengineering, is the intervention in the climate system to counter climate change by rectifying the radiative forcing imbalance~\cite{vaughan2011review,soldatenko2017weather}. 
Climate engineering is divided into two broad categories: \emph{Carbon Dioxide Removal} (CDR) and \emph{Solar Radiation Management} (SRM).
CDR affects the link from emissions to concentrations, SRM aims to break the link from concentrations to temperatures.

CDR aims to remove $\text{CO}_2$ from the atmosphere by creating or enhancing carbon sinks.
CDR encompasses a variety of solutions, e.g., land carbon sink enhancement via afforestation and reforestation, and ocean carbon sink enhancement through increasing ocean alkalinity, and iron fertilization.
The effect of CDR options decays over time, because of the response of the ocean and land carbon reservoirs to atmospheric perturbations, and if the carbon storage is not permanent. 

SRM aims to decrease the radiative forcing 
by reducing the amount of solar radiation absorbed.
SRM could be achieved through sunshades in space, stratospheric aerosols (i.e., injection of sulphate aerosols into the lower stratosphere emulating large volcanic eruptions), enhancing cloud albedo (e.g., increasing the reflectivity of low level marine stratiform clouds), or surface albedo (e.g., increasing grassland and cropland albedo).
SRM methods have common issues; in particular, reducing incoming solar radiation does not ameliorate ocean acidification caused by rising atmospheric $\text{CO}_2$.

It is worth pointing out that climate engineering and its effect on the climate system are typically not explicitly accounted for in the mentioned IAMs.
Indeed, CDR technologies are included only in a few attempts, e.g., see (in GCAM) \cite{morrow2023gcam} or (in DICE) \cite{rickels2018integrated}, while SRM is not yet, even though it would clearly influence the radiative forcing.
Experimental and synthetic data from detailed climate simulators, see \cite{kravitz2016geoengineering}, can be used to learn to explicitly model the effect of geoengineering on the climate system, e.g., the effect of SRM technologies on radiative forcing.



\section{A feedback control framework}
\label{sec:feedback}

This paper suggest to adopt a feedback control approach in the HESs study. We proposes a control framework that, using measurements from the HES,  output control setpoints performing, e.g., climate change adaptation and mitigation actions. The framework is constituted by:
\begin{enumerate}
    \item An outer and slower control loop whose goal is to identify  policies, i.e., functions that, given the state of the HES, prescribe the optimal action. 
    \item An inner and faster control loop whose goal is to actuate the setpoints provided by the aforesaid policies.
\end{enumerate}
To formalize the proposed framework, following a control engineering perspective, the states of the HES and the control inputs need to be identified. Let the vectors $\bx, \bu$ collect the states and the inputs, respectively\footnote{For simplicity, we assume that the states are directly measured, though a more general framework where states are not accessible can be handled by adding a state estimator, e.g., a Kalman filter.}.
The state $\bx$, belonging to a state set $\mcX$, comprises the variables needed to describe and predict the system behavior with reasonable accuracy.
States typically chosen in the literature include:
\begin{enumerate}
\item \emph{Climate variables}, e.g., $\text{CO}_2$ concentrations in the atmosphere, carbon stock in the soil and oceans, and the average temperature.
\item \emph{Socio-economic variables}, e.g., the overall population, the world gross product, the capital stock, the Renewable energy knowledge stock.
\end{enumerate}
Control inputs $\bu$, belonging to a state set $\mcU$, represent the variables that can be directly regulated to steer the HES toward targeted values and may include:
\begin{enumerate}
\item \emph{Climate inputs}, e.g., $\text{CO}_2$ amount captured and stored by CDR technologies, the amount of sprayed particles in the stratosphere, and afforestation goals.
\item \emph{Socio-economic inputs}, e.g., a carbon tax, the capital invested in the production processes (\cite{barrage2023policies}) or in renewable energy, the gross world product growth rate~\cite{Strnad2019DRL}.
\end{enumerate}
Formally, the HES can be modeled by a vector function 
$$\bff:\mcX \times \mcU \rightarrow \mcX, (\bx,\bu)\mapsto \bff(\bx,\bu).$$

\begin{figure}[tb]
\centering
\includegraphics[width= 0.95\columnwidth]{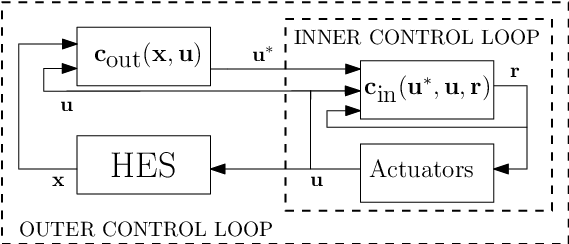}
\caption{Block scheme representation of the proposed control framework.}
    \label{fig:ClimSys}
\end{figure}

Figure~\ref{fig:ClimSys} graphically depicts the proposed control architecture.
The controller governing the outer control loop, denoted as $\bc_{\text{out}}$, computes and updates the HES control setpoint $\bu^*$  based on the HES state and on the implemented input $\bu$.
The inner loop controller, denoted as $\bc_{\text{in}}$, ensures that $\bu^*$ is actually applied to the HES.
The two control loops might act on two different time scales: 
\begin{enumerate}
\item the outer loop on a slower one, since HESs evolve on several months/a few years time intervals.
\item the inner loop on a faster one, since, ideally, $\bu^*$ should be quickly implemented.
\end{enumerate}
Notably, control theory gives the tools to: (i) handle the time scale separation, e.g., by exploiting singular perturbation theory; (ii) provide formal conditions for the stability of the closed loop system; (iii) provide performance guarantees.

\section{Optimization of HES}
\label{sec:opt}






The goal is to compute the action/intervention $\bu^*$ for the HES solving the multi-stage optimization problem
\begin{subequations}\label{eq:OPT}
	\begin{align}
	\bu^*:=\arg\min_{\bu \in \mcU}\ &  ~\sum_{t = 1}^{T} J(\bx(t),\bu(t)) \label{eq:OPT:cost}\\
	\mathrm{s.t.}\  & ~\bx(t+1) = \bff(\bx(t),\bu(t)) \label{eq:OPT:eq}\\
	\  & ~\bg(\bx(t),\bu(t)) \leq 0 \label{eq:OPT:ineq}
	\end{align} 
\end{subequations}
where the objective function has the form
$$J: \mcX \times \mcU \rightarrow \mathbb R, \quad (\bx,\bu) \mapsto J(\bx,\bu).$$
The constraint~\eqref{eq:OPT:eq} captures the HES behavior.
The inequality constraint~\eqref{eq:OPT:ineq}, where 
$$\bg : \mcX \times \mcU \rightarrow \mathbb R^{N}, (\bx,\bu) \mapsto g(\bx,\bu) $$
maps the HES states and control inputs into $N$ constraints that describe desirable system properties.
For instance, $\bg$ can capture effects such as the global mean temperature increase or precipitation changes, effects of climate change on renewable energy generations, access to food and water.

State-of-the-art IAMs  obtain $\bu^*$ by \emph{numerically} solving problems like~\eqref{eq:OPT}.
Because of unavoidable errors and uncertainties in the peculiar HES models they use, different IAMs arrive at strikingly different solutions.
Hence, the solution $\bu^*$ is not necessarily the \emph{true} optimal setpoint and depends on the used HES model $\bff$. 
Table~\ref{table:1} reports the generation mixes computed by three IAMs for two \emph{Shared Socioeconomic Pathways} (SSPs)\footnote{SSPs are 
scenarios of projected socioeconomic global changes defined in the IPCC Sixth Assessment Report on climate change. SSPs are used to derive GHG emissions under 
different climate policies. 
In SSP1, consumption is oriented toward low material growth and lower resource and energy intensity with a resulting low global average temperature rise. 
SSP2 represents a scenario where social, economic, and technological trends do not shift markedly from historical patterns. Even though environmental systems degrade, overall the intensity of resource and energy use declines.}.
Structural differences between models lead to energy mixes that differ both in the adopted technologies and in the amount of produced energy~\cite{pindyck2017use}.
For these reasons, there are even scientists claiming the very extreme statement that IAMs are \emph{useless} and create a perception of knowledge and precision without scientific legitimacy, 
 see~\cite{pindyck2013climate,pindyck2017use}.

\begin{table}[t]
\centering
\begin{tabular}{|c||c|| c| c| c| } 
\hline \hline
\multicolumn{2}{|c||}{}  & GCAM & IMAGE & WITCH \\
\hline
\multirow{ 7}{*}{SSP1} & Coal & 173.5 & 116.1 & 83.5 \\
& Oil & 180.9 & 54.5 & 164.1 \\
& Gas & 162.3 & 216.9 & 140.6 \\
& Nuclear & 2.3 & 0 & 16.2 \\
& Biomass & 94.5 & 137.4 & 129.2 \\
& Renewables & 161.9 & 181.2 & 196.7 \\
& Total Energy & 775.4 & 706.1 & 730.1 \\
\hline \hline 
\multirow{ 7}{*}{SSP2} & Coal & 431.2 & 509.8 & 234.7 \\
& Oil & 349.8 & 180.5 & 360.1 \\
& Gas & 209.9 & 215.4 & 301.6 \\
& Nuclear & 61.9 & 11.9 & 42.2 \\
& Biomass & 109.1 & 116.5 & 136.8 \\
& Renewables & 87.2 & 123.4 & 202.9 \\
& Total Energy & 1249.1 & 1145.6 & 1278.3 \\
\hline \hline 
\end{tabular}
\vspace{2mm}
\caption{Energy mixes proposed by three IAMs in [EJ], see~\cite{riahi2017shared}.}
\label{table:1}
\end{table}

\begin{figure}[tb]
\centering
\includegraphics[width= 0.95\columnwidth]{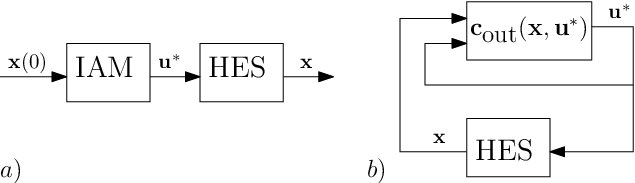}
\caption{a) IAMs act in an open loop fashion. b) The closed loop approach.}
    \label{fig:CSloop}
\end{figure}

From a system and control perspective, IAMs act in an \emph{open loop} fashion: That is, the control input $\bu^*$ is computed by solving problem~\eqref{eq:OPT} relying only on the considered HES model and on his initial conditions, see Figure~\ref{fig:CSloop}$a)$. The IAMs substantially different results can then be explained by the well-known fact that open loop strategies are very sensitive to model inaccuracies and uncertainties.

Conversely, problem~\eqref{eq:OPT} can be solved in a \emph{closed loop} fashion, e.g., by utilizing tools from Model Predictive Control (MPC)~\cite{alessio2009survey}, or feedback-based optimization~\cite{Cavraro2022Feedback}, see Figure~\ref{fig:CSloop}$b)$. The HES dynamically evolves toward $\bu^*$ following the input computed by the control function  
$$\bc_{\text{out}}: \mathcal \mcX \times \mcU \rightarrow \mathcal U, \quad (\bx,\bu) \mapsto \bc_{\text{out}}(\bx,\bu).$$
which updates the control input from measurements of the actual state and inputs.
%
%
%
We envision this approach providing benefits from the fact that closed loop strategies:
\begin{enumerate}
    \item have well-established properties of disturbance rejection: they actively compensate model uncertainties, inaccuracies and disturbances using measurements.
    \item update and adjust the control input using the information provided by the measurements until they reach (asymptotically) the desired steady state value.
    \item can be tuned to obtain desired transitorial behavior, e.g., in terms of setting time or overshoot. 
\end{enumerate}
For instance, adopting an MPC approach, problem~\eqref{eq:OPT} can be solved to obtain a sequence $\bu^*(1),\bu^*(2),\dots,\bu^*(T)$. Then, only the first
input $\bu^*(1)$ is applied to the HES and the optimization problem~\eqref{eq:OPT} is solved again based on the applied control setpoint and the new measured state.

Because the variables of interest, e.g., the average temperature, evolve slowly, the outer loop controller acts and adjusts $\bu^*$ on a time scale of several months/a few years.



\section{Control of Actuators}
\label{sec:act}

\emph{Actuators} are agents interacting with the HES that are  responsible for implementing the optimal setpoints computed in the outer control loop.
Actuator is used here as an abstract word that in practice has different form depending on the system where is acting. Examples includes
\begin{enumerate}
\item in \emph{climate systems} carbon capture and storage facilities, plants for ocean fertilization.
\item in \emph{socio-economic systems}, policy and decision makers setting a carbon tax or funding renewables.
\end{enumerate}
We formally define the actuators map as the function
$$\ba: \mathcal R \rightarrow \mathcal U, \quad \br \mapsto \ba(\br) $$
that outputs the control inputs for the HES, where $\mathcal R$ is the set of admissible reference signals. 
The goal is to create a sequence of inputs $\{\bu(t)\}$ converging to the desired value $\bu^*$, see Fig.~\ref{fig:ClimSys}. 
For example, consider an SRM application. The vector $\bu^*$ will contain the desired albedo and the vector $\br$ will contain the instructions for the actuators, e.g., the location and the amount of particles sprayed by aircraft in the stratosphere.
The work~\cite{kravitz2016geoengineering} proposes a proportional-integral (PI) controller for an SRM application, where the goal is to regulate the albedo by means of stratospheric aerosol injections. From measurements of temperature, the PI scheme ensures the implementation of the desired solar reduction.
Other techniques could be applied to manage actuators, like reinforcement learning~\cite{zhang2022application}.

Feedback-based optimization~\cite{Cavraro2022Feedback} looks a promising tool for controlling actuators and for finding the best $\br$ implementing $\bu^*$. First, cast an optimization problem like
\begin{subequations}\label{eq:F&O}
	\begin{align}
	\min_{\br \in \mathcal R}\ &  ~\phi(\bu^*,\bu,\br) \label{eq:F&O:cost}\\
	\mathrm{s.t.}\  & ~\bu = \ba(\br) \label{eq:F&O:eq}
	\end{align} 
\end{subequations}
A meaningful objective function in~\eqref{eq:F&O:cost}, $\phi:\mcU \times \mcU \times \mathcal{R} \rightarrow \mathbb R$, could penalize a weighted sum of the difference between $\bu^*$ and the actuator output and the cost of $\br$, e.g.,
\begin{equation}
\phi(\bu^*,\bu,\br) = \|\bu^*-\bu\|^2 + \alpha \|\br\|^2
\label{eq:cost_inner}
\end{equation}
with $\alpha$ being a positive constant. By plugging~\eqref{eq:F&O:eq} into~\eqref{eq:F&O:cost}, problem~\eqref{eq:F&O} becomes
$$\min_{\br \in \mathcal R} \ \phi(\bu^*,\ba(\br),\br).$$
Instead of solving~\eqref{eq:F&O} numerically and apply directly its solution, feedback-based optimization adopts an iterative approach,
e.g., the projected gradient descent algorithm
\begin{equation}
\br(t+1) = \Big[ \br(t) - \epsilon \frac{\partial \phi(\bu^*,\bu(t),\br(t))}{\partial \br} \Big]_{\mathcal R}
\label{eq:PGD}
\end{equation}
where $[\cdot]_{\mathcal{R}}$ represents the projection over $\mathcal R$ and $\epsilon$ is a suitable constant that guarantees the convergence of~\eqref{eq:PGD}. 
The updated reference value $\br(t+1)$ is every time applied to the system.
The computation of $\partial \phi$ usually exploits measurements that represent the feedback from the system, i.e., how the system reacted to the new reference, and reduce the effect of modeling errors on $\ba$.  
The inner loop controller can then in general be expressed as the function
$$\bc_{\text{in}}: \mathcal U \times \mathcal U \times \mathcal R \rightarrow \mathcal R, \quad (\bu^*,\bu,\br) \mapsto \bc_{\text{in}}(\bu^*,\bu,\br).$$
Heed that other optimization algorithms have seen a feedback implementation, e.g., primal-dual methods, dual ascent, ADMM.  
The iterative approach ensure that the sequence $\ba(\br(1)),\ba(\br(2)),\dots$ converges (close) to $\bu^*$.

%




\section{Application Examples}

This Section presents a few applications where the proposed control framework could be applied. In each example, we identify the ingredients, e.g., the state and control vectors $\bx$ and $\bu$, the HES model $\bff(\bx,\bu)$, the cost $J(\bx, \bu)$ that can be plugged into~\eqref{eq:OPT} and~\eqref{eq:F&O} to formulate the problem of interest.

\subsection{SRM via Stratospheric Aerosol Injection (SAI)}
SAI seeks to reproduce the cooling effect of volcanic eruptions by injecting sulfates into the stratosphere and forming a reflective aerosol. 
Denote by $T(\psi)$ the \emph{zonal-mean temperature} at latitude $\psi$.
SAI aims at shaping $T(\psi)$ to meet climate objectives, e.g., keeping global warming below $2^\circ$ C.

Recent studies propose to achieve different zonal average stratospheric sulfate distributions for regulating the global mean temperature $T_0$, the interhemispheric temperature gradient $T_1$, and the equator-to-pole temperature gradient $T_2$. 
This is to decrease the average temperature and preserve other important climate features like precipitation patterns.
$T_0, T_1, T_2$ can be computed as the projection of $T(\psi)$ into the three Legendre polynomials~\cite{kravitz2016geoengineering}
$$
L_0(\psi) = 1, \quad
L_1(\psi) = \sin \psi, \quad
L_2(\psi) = 1/2(3 \sin^2\psi - 1).
$$
The Legendre polynomials are also used as a basis to describe zonal average stratospheric
sulfate distributions which affect the radiative forcing~\cite{ban2010geoengineering}.
Denote by $D(\psi)$ the Aerosol Optical Depths (AODs) at the latitude $\psi$.
The projections of $D(\psi)$ on the Legendre polynomials are denoted as $D_0, D_1, D_2$, respectively.
Acting on $D_0, D_1, D_2$ produces controllable changes in $T_0, T_1, T_2$~\cite{macmartin2017climate}.
Aerosols can be injected at high altitude (roughly 5 km above the tropopause) at just four locations, i.e., $30^\circ$ N, $15^\circ$ N, $15^\circ$ S, $30^\circ$ S in latitude along $180^\circ$ longitude, to obtain an approximation of stratospheric sulfate distributions expressed as any combination of the above polynomials.
For example, different combinations of injections in these four locations can produce patterns of aerosol optical depths that are constant, linear, and quadratic in the latitude~\cite{macmartin2017climate,kravitz2017first}.

The goal is to  affect the radiative forcing via SAIs to counter climate change and support GHG emissions cut policies.  
The HES here is a coupled atmosphere-ocean general circulation model capable of representing the formation and interactions of stratospheric sulfate aerosol from source gases and the radiative forcing, e.g., the fully-coupled Community Earth System Model (CESM)~\cite{kravitz2017first}. The state $\bx$ comprises the variables appearing in the HES equations.
Given that any zonal-mean temperature can be approximated as a combination of the Legendre polynomials 
$$T(\psi) = T_0 L_0(\psi) + T_1 L_1(\psi) + T_2 L_2(\psi).$$
and after defining the control input 
$\bu = \begin{bmatrix}
T_0 & T_1 & T_2    
\end{bmatrix}^\top$, an interesting outer optimization problem aims to find the parameters for the best target zonal-mean temperature $T^*(\psi)$.
A meaningful metric could be penalizing the control effort
$$J(\bx(t),\bu(t)) = \| \bu(t) \|.$$
to minimize human intervention. The function $\bg(\bx,\bu)$ models how $\bu$ influences the quantities of interest, e.g., temperatures, temperature gradients, and precipitation patterns.

Actuators are aircraft injecting sulfates into the stratosphere. The goal of the inner control loop is to have the optimal control input $\bu^*$ implemented using SAI. The vector $\br$ collects then the amount of particles to inject in the aforementioned four locations. Assume that the function 
$\boldsymbol \varphi: \mathcal R \rightarrow \mathcal D, \br\mapsto \boldsymbol 
 \varphi(\br)$
maps injections into the AOD components $D_0, D_1, D_2$, with
$$D(\psi) = D_0 L_0(\psi) + D_1 L_1(\psi) + D_2 L_2(\psi).$$
where $\mathcal D$ is the set of feasible $D_0, D_1, D_2$. Also, let $\xi$ be the function modeling how AODs affect the temperature
$$\boldsymbol \xi: \mathcal D \rightarrow \mcU, 
\begin{bmatrix}
D_0 & D_1 & D_2       
\end{bmatrix}^\top \mapsto \boldsymbol \xi(\begin{bmatrix}
D_0 & D_1 & D_2       
\end{bmatrix}^\top)$$ 
The functions $\boldsymbol \varphi$ and $\boldsymbol \xi$ can be learned from simulations~\cite{kravitz2017first}. 
The actuator map is given then by the composition
$$\ba(\br) = \boldsymbol \xi (\boldsymbol \varphi (\br))$$
The system feedback is captured by measurements of temperatures and AOD, which can be performed using satellites.
For instance, observations from the Moderate Resolution Imaging Spectroradiometer (MODIS) on NASA's Terra satellite are already used to monitor aerosol amounts around the world.

A feedback-based optimization algorithm like~\eqref{eq:PGD} could then be developed to solve problem~\eqref{eq:F&O} adopting the cost~\eqref{eq:cost_inner}.

\subsection{Developing sustainable economic pathways.}
IAMs are extensively used to evaluate the climatic and social impacts of economic growth scenarios. As mentioned earlier, different modeling choices across IAMs lead to different solutions.
The differences can be alleviated by solving~\eqref{eq:OPT} using feedback control rather than numerically.
For example, consider the model used in DICE~\cite{barrage2023policies}.
Society preferences are modeled by a social welfare function that ranks different paths of consumption.
The DICE model assumes that economic and climate policies should be designed to optimize the flow of consumption over time.
The HES state $\bx$ encompasses:
\begin{enumerate}
    \item \emph{climate variables}: Earth surface and oceans temperature, carbon stock in the land and the oceans, $\text{CO}_2$  atmospheric concentration and emissions;
    \item  \emph{socio-economic variables}: the world population (denoted as $\pi(t)$) and  gross product, the productivity, the per-capita consumptions, the damage of climate change.    
\end{enumerate}
The HES~\eqref{eq:OPT:eq} comprises twenty one nonlinear equations~\cite{barrage2023policies}; 
the control $\bu$ includes the emission control rate, which influences the cost of emissions reductions, and capital consumption (denoted as $\gamma(t)$) and investments.
The outer optimization problem objective function models the social welfare 
\begin{equation*}
J(\bx(t),\bu(t)) = \frac{\pi(t) \left( \frac{\gamma(t)}{\pi(t)}\right)^{1-\eta}}{1-\eta} \Psi(t).
\end{equation*}
$J(\bx(t),\bu(t))$ represents the discounted sum of the per capita consumption, where $\Psi(t)$ is the discount factor at time $t$ and $\eta$ models the marginal utility of consumption.
The welfare function is increasing in the per capita consumption of each generation, with diminishing marginal utility of consumption. 
The constraint~\eqref{eq:OPT:ineq} captures desirable outcomes like a threshold on the average temperature change.

Actuators here are policymakers. The outer control recommends the best control actions $\bu^*$  and updates $\bu^*$ based on how the HES reacts to the applied control setpoint. 
The actuator model $\ba$ can be obtained using tools from implementation theory~\cite{pal1997beyond}.

\subsection{Marine cloud brightening for coral reef preservation}
In the former examples, the control $\bu^*$ consists of actions that should be implemented on a \emph{global} scale. This may raise concern about the implementability of such control actions, as different Countries have different priorities or values. Conversely, here we present a \emph{local} application.
Scientists are testing if marine cloud brightening (MCB) could be effective in cooling the Great Barrier Reef to preserve the coral population~\cite{tollefson2021can}.
The coral reefs and ecosystems are indeed endangered because of a combination of increased ocean temperatures and acidification. 
MCB involves spraying sea salt aerosols into marine stratocumulus to increase cloud reflectivity. 
The proposed control framework can be used to plan and actuate MCB interventions. 

The HES model comprises climatic equations modeling the dynamics in the atmosphere, ocean, and land in the region of interest. A recent work adopted the fully-coupled Community Earth System Model (CESM)~\cite{macmartin2023transboundary}. 
The state $\bx$ includes the climate variables appearing in the HES equations, e.g., the atmospheric temperatures.
The control input $\bu$ represents the surface albedo.

The outer loop finds $\bu^*$ by solving an optimization problem whose cost penalizes the deviation from a desired water temperature $\tau$. The water temperature, denoted as $\rho$, depends on the HES state and the albedo reduction. After modeling the water temperature as a function 
$$\rho: \mcX \times \mcU \rightarrow \mathbb R, \quad (\bx,\bu) \mapsto \rho(\bx,\bu)$$
a cost function that can be used in~\eqref{eq:OPT} is
$$J(\bx(t),\bu(t)) = \| \tau - \rho(\bx(t),\bu(t)) \|.$$
The optimal albedo reduction obtained by solving~\eqref{eq:OPT} will be physically implemented. Actuators will be boats endowed with turbines that blow a mist of seawater that ascends into the sky whose model is the function $\ba(\br)$.
The reference signal $\br$  regulates the functioning of the turbines to obtain the desired albedo reduction.
The system feedback consists of aerosol measurements possibly taken by drones, satellites, or other boats~\cite{tollefson2021can}. 

\section{A numerical example}

Here, we provide some numerical experiments to illustrate the benefit of adopting a feedback approach when the model parameters are affected by uncertainty in the context of evaluate and plan economic pathways for climate change mitigation.
For our tests, we considered the AYS model, which is a low-complexity model incorporating climate change, welfare
growth and energy transformation~\cite{Strnad2019DRL}.

The system's state comprises three variables.
$A$ [GtC] is the atmospheric carbon stock in  excess w.r.t. a pre-industrial level;
$Y$ [\$/yr] is the gross world product (GWP) and is an indicator of the wealth of the society;
$S$ [GJ] is the renewable energy knowledge stock. It  models the know-how for the production of renewable energy and is identified  with the past cumulative production of renewables. Hence, 
$\bx^\top =
\begin{bmatrix}
A & Y & S  
\end{bmatrix}^\top
$.
The control inputs are the growth index change $\beta$ [\%$\ \text{yr}^{-1}$] which models the GWP's growth, and the break-even knowledge level $\sigma$ [GJ]  at which renewable and fossil extraction costs become equal which can be changed by subsidizing renewables or introducing a carbon tax. That is,
$\bu^\top =
\begin{bmatrix}
\beta & \sigma   
\end{bmatrix}^\top
$. 
The system obeys
%
%
\begin{equation}
\dot \bx = \bff(\bx,\bu), \quad \bff(\bx,\bu) = 
\begin{bmatrix}
\frac{\sigma^\rho}{\sigma^\rho + S^\rho} \frac{Y}{\epsilon \phi} - \frac{A}{\tau_A}\\
(\beta - \theta A) Y \\
\frac{S^\rho}{\sigma^\rho + S^\rho} \frac{Y}{\epsilon} - \frac{S}{\tau_S} 
\end{bmatrix}
\label{eq:AYS_ODE}
\end{equation}
Equation~\eqref{eq:AYS_ODE} represents an ODE parametrized in
the climate change damage coefficient $\theta \; [\text{yr}^{-1} \text{GtC}^{-1}]$,
the energy efficiency $\epsilon \;[\$ \, \text{GJ}^{-1}]$,
the fossil fuel combustion efficiency $\phi \; [ \text{GJ GtC}^{-1}]$,
the time constants $\tau_A = 50$ yr and $\tau_S = 50$ yr.  
The initial state and control inputs are
$$A_0 = 840 \text{ GtC}, \, Y_0 = 7 \cdot 10^{13} \$ \text{ yr}^{-1}, \, S_0 = 5 \cdot 10^{11} \text{ GJ}$$
$$\beta_0 = 3 \% \text{ yr}^{-1} \quad \sigma_0 = 5 \cdot 10^{12} \text{ GJ},$$
The system's feedback consists of state measurements obtained once every two years.

We assume that the goal is to track a desired trajectory for the level of atmospheric $\text{CO}_2$. 
To that aim, we cast an optimization problem in which the objective function is
\begin{equation*} J(\bx,\bu) = \frac{A^2}{\lambda} + \frac{(\sigma - \sigma_0)^2}{\mu} + \frac{(\beta - \beta_0)^2}{\nu} + \bu^\top 
\begin{bmatrix}
w_\beta & 0\\
0 & w_\sigma    
\end{bmatrix}
\bu
\label{eq:AYS_cost}
\end{equation*}
with $\lambda, \mu, \nu, w_\beta, w_\sigma$ positive parameters.
A discretization of~\eqref{eq:AYS_ODE} provides the equality~\eqref{eq:OPT:eq}.
The operational constraints~\eqref{eq:OPT:ineq} are replaced by the equations 
\begin{subequations}
\begin{align}
& A,Y,S,\sigma \geq 0\label{eq:in_a}\\
& A \leq 345, \quad Y \geq 4 \cdot 10^{13} \label{eq:in_b}
\end{align}
\end{subequations}
Inequality~\eqref{eq:in_a} captures hard constraints on states and control inputs. Further,~\eqref{eq:in_b} describes a target subspace ensuring a certain level of welfare and in which the level of atmospheric $\text{CO}_2$ is reduced. 
To simulate the fact that uncertainties affect the model, we assume that the available model parameters are 
$\theta = 8.57 \cdot 10^{-5}, \epsilon = 147, \phi = 4.7 \cdot 10^{10}$
whereas their \emph{true} value can deviate up to 20\%.

A numerical (open loop) solution to the problem provides the control setpoints to track the desired trajectory of $A$. 
However, due to the parameter errors, applying these control inputs results in higher levels of atmospheric $\text{CO}_2$. 
Adopting instead a closed-loop approach (precisely, implementing an  MPC algorithm), we can track the desired trajectory of $A$ up to an initial transitorial phase. 

\begin{figure}[tb]
\centering
\includegraphics[width= 0.99\columnwidth]{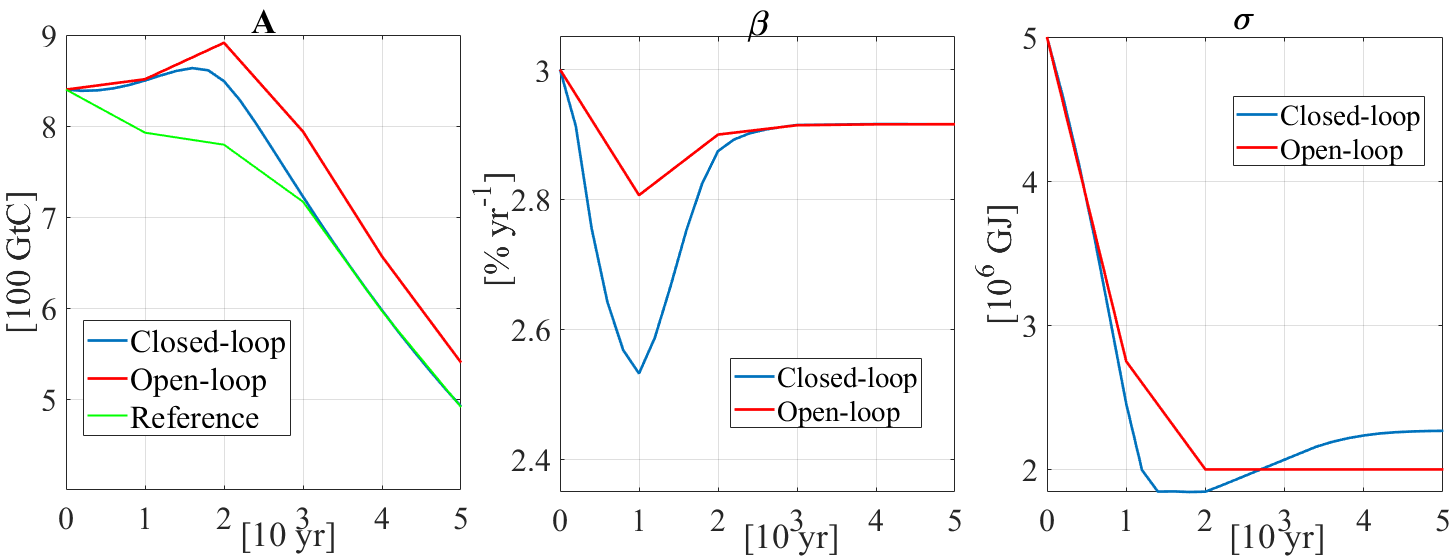}
\caption{Left panel: the excess atmospheric carbon stock, Center panel: the GWP's growth rate. Right panel: the break-even knowledge level.}
    \label{fig:sym}
\end{figure}

The results are reported in Figure~\ref{fig:sym}. 
We can observe that the closed-loop control input evolutions differ significantly from the open-loop one.
Indeed, the closed-loop solution prescribes control actions that are more drastic than the open-loop one, which would yield a degraded trajectory of $A$ roughly delayed by about five years w.r.t. the reference signal.

\section{CONCLUSIONS}
\label{sec:conclusions}

We argue that control theory is an effective tool for HESs, e.g., to tackle the unprecedented challenge of climate change.
Even though HES science already encompasses aspects from several disciplines, e.g., climate science, economics, and social science, our goal is to add a (control) engineering perspective in a field that has been historically mainly a prerogative of economists or Earth and climate scientists. 
Feedback control allows us to find the optimal HES inputs mitigating the effects of modeling uncertainties and errors.
This paper proposes a control framework that features an outer loop and an inner control loop. The outer loop iteratively updates the control inputs needed to steer the HES to desirable configurations; the inner control loop ensures that the control setpoints computed in the outer loop are implemented by actuators. Possible applications are also presented.



\bibliographystyle{ieeetr}
\bibliography{Bibliography}

\end{document}